\documentclass[journal]{IEEEtran}
\usepackage{cite}
\usepackage{amsmath,amssymb,amsfonts}
\usepackage{algorithmic}
\usepackage{textcomp}
\def\BibTeX{{\rm B\kern-.05em{\sc i\kern-.025em b}\kern-.08em
    T\kern-.1667em\lower.7ex\hbox{E}\kern-.125emX}}

\usepackage{textcomp}
\usepackage{csvsimple} 
\usepackage{dblfloatfix}  
\usepackage{hyperref}
\usepackage{filecontents,url}
\usepackage{blindtext}
\usepackage{float}
\usepackage{subfig}
\usepackage[pdftex]{graphicx}
\usepackage{tikz}

\definecolor{accessblue}{RGB}{0,105,154}

\newcommand{\rev}[1]{\textcolor{black}{#1}}

\DeclareMathOperator*{\argmin}{arg\,min}

\DeclareMathOperator{\Network}{\ensuremath{\Lambda}}

\newcommand{\data}{\ensuremath{y}}
\newcommand{\noise}{\ensuremath{\delta\data}}

\begin{document}

\title{Unsupervised denoising for sparse multi-spectral computed tomography}

\author{Satu I. Inkinen, Mikael A. K. Brix, Miika T. Nieminen, Simon Arridge, Andreas Hauptmann,
\thanks{This work was supported by Academy of Finland (project: 316899, 336796, 334817, 338408) and the CMIC-EPSRC platform grant (EP/M020533/1).}
\thanks{Satu I. Inkinen, Mikael A. K. Brix and Miika T. Nieminen are with the Research Unit of Medical Imaging, Physics and Technology, University of Oulu, Aapistie 5A, 90220, Oulu, Finland (e-mail: satu.inkinen@oulu.fi; mikael.juntunen@oulu.fi; miika.nieminen@oulu.fi ).}
\thanks{Simon Arridge is with the Department of Computer Science, University College London, WC1E 6BT London, U.K. (e-mail: s.arridge@cs.ucl.ac.uk).}
\thanks{Andreas Hauptmann is with the Research Unit of Mathematical Sciences,
University of Oulu, Pentti Kaiteran katu 1, 90570, Oulu, Finland and with the Department of Computer Science, University College London, WC1E 6BT London, U.K (e-mail: andreas.hauptmann@oulu.fi).}
}

\maketitle

\begin{abstract}
Multi-energy computed tomography (CT) with photon counting detectors (PCDs) enables spectral imaging as PCDs can assign the incoming photons to specific energy channels. However, PCDs with many spectral channels drastically increase the computational complexity of the CT reconstruction, and bespoke reconstruction algorithms need fine-tuning to varying noise statistics. \rev{Especially if many projections are taken, a large amount of data has to be collected and stored.  Sparse view CT is one solution for data reduction. However, these issues are especially exacerbated when sparse imaging scenarios are encountered due to a significant reduction in photon counts.}  In this work, we investigate the suitability of learning-based improvements to the challenging task of obtaining high-quality reconstructions from sparse measurements for a 64-channel PCD-CT. In particular, to overcome missing reference data for the training procedure, we propose an unsupervised denoising and artefact removal approach by exploiting different filter functions in the reconstruction and an explicit coupling of spectral channels with the nuclear norm. Performance is assessed on both simulated synthetic data and the openly available experimental Multi-Spectral Imaging via Computed Tomography (MUSIC) dataset. We compared the quality of our unsupervised method to iterative total nuclear variation regularized reconstructions and a supervised denoiser trained with reference data. We show that improved reconstruction quality can be achieved with flexibility on noise statistics and effective suppression of streaking artefacts when using unsupervised denoising with spectral coupling. 

\end{abstract}

\begin{IEEEkeywords}
Computed tomography, deep learning, unsupervised methods, image reconstruction, sparse angles, spectral imaging 
\end{IEEEkeywords}

\section{Introduction}
\label{sec:introduction}
\IEEEPARstart{S}{pectral}  computed tomography (CT) has recently emerged as a rapidly growing imaging modality as photon counting detector (PCD) technology, in particular, has made significant advances in past years \cite{Leng2019, Dudak2020}. Spectral CT provides several advantages over conventional single energy CT such as enhanced material discrimination ability, quantitative imaging, and energy selective imaging \cite{Lee2017a, Juntunen2020}. In the medical imaging domain,  spectral CT with PCDs has been successfully applied, for example, in the characterization of gout \cite{Huber2020} and contrast-enhanced imaging of the kidneys \cite{Pourmorteza2016}. In the security industry, spectral imaging has enabled automated detection and characterization of hazardous materials e.g. explosives and weapons \cite{Mouton2015a, Jumanazarov2020}. Furthermore, multi-bin PCDs having more than 64 energy channels have enabled spectroscopic CT imaging, which can be exploited in material classification tasks \cite{Busi2019, Jumanazarov2020}. For example, a material classification method was applied in estimating relative electron density 
and effective atomic number 
maps based on the corrected linear attenuation coefficient (LAC) data from spectral channel reconstructions \cite{Busi2019, Jumanazarov2020}.

\rev{If the object is imaged with thousands of projections in multi-channel spectral CT, the amount of acquired data may become problematic.  Especially, if the projection images are acquired in 3D. With sparse spectral CT, the number of projections is reduced. The sparse-view CT has been mainly studied for single-energy CT \cite{Kisner2012,  Zhang2018a}, however, due to the increased data with the additional energy channel dimension in spectral CT, it is a useful method for data reduction. For example in airport security applications where rapid data acquisition and processing is needed, effective spectral sparse CT could be useful.}

As only a limited number of photons arrive in the spectral channels, the acquired projection data is inherently noisy. Therefore, dedicated noise-robust spectral reconstruction methods have been developed for spectral PCD-CT  \cite{Siltanen2020,Semerci2014,Zhang2017}. These methods make use of iterative reconstruction algorithms and cover joint regularization methods \cite{Siltanen2020}, tensor-based nuclear norm regularization \cite{Semerci2014,rigie2015joint}, and dictionary learning \cite{Zhang2017} approaches. However, the above studies focus on PCDs with a maximum of eight energy channels, and as the number of energy channels increases, the computation time of the iterative methods will become a challenge – especially when the spectral projection data is acquired in 3-D. Therefore, alternative approaches have to be sought and developed to speed up the channel-wise reconstructions of spectral multi-bin CT. 
Prior studies utilizing multi-bin (over 64 energy channels) data have mainly used filtered back projection or iterative reconstruction methods such as ART or SIRT after reduction of the energy channels \cite{Kehl2018, Jumanazarov2020}, where the focus has mainly been on the material differentiation  aspect. Kehl \textit{et al.}  \cite{Kehl2018} compared different unsupervised segmentation approaches for the openly available spectral CT dataset (MUSIC), also used in this study, collected from a 128-bin PCD utilizing iterative ART-TV reconstruction. 

Recently, data-driven approaches have been developed for single energy CT reconstruction \cite{Kang2017,Jin2017,Adler2017,Chen2018,Shan2018,adler2018learned} to improve image quality with a considerable speed-up in reconstruction times compared to iterative approaches after the training is performed. 
For single energy CT reconstruction, several different learning approaches have been developed. The most common can be roughly divided into two categories: post-processing \cite{Kang2017,Jin2017,Shan2018,pelt2018mixed} and learned iterative networks \cite{Adler2017, Chen2018,adler2018learned,hauptmann2020multi}. 
Post-processing networks primarily focus on removing noise and artefacts from an initial reconstruction, obtained, for instance, by filtered back projection (FBP), and are especially time-efficient. Learned iterative approaches, on the other hand, are more time-consuming due to the repeated application of the projection operator. 
Nevertheless, regardless of the chosen network architecture, both approaches provide significant improvements in image quality over conventional iterative variational methods for single energy CT. 
For multi-spectral CT, most data-driven approaches address the material decomposition problem, by training a network to perform the decomposition either in data space prior to reconstruction \cite{zimmerman2015experimental,touch2016neural,Abascal2020ISBI,abascal2021material} or afterwards \cite{clark2018multi}. In a recent study, Mustafa et al. \cite{mustafa2020sparse} have investigated the use of a post-processing network for the MUSIC data trained in a supervised manner. Their post-processing network improved the image quality of the FBP images and achieved higher image quality when compared to the iterative ART-TV reconstruction \cite{mustafa2020sparse}.

In particular, the majority of learning-based reconstructions are trained supervised, which means that a relatively large set of ground-truth reconstructions is needed. This is especially problematic for applications where no densely sampled measurement is available, and hence one would need to resort to either synthetically generated simulated data, as in \cite{mustafa2020sparse}, to train the networks or use gold-standard reference reconstructions from the low-quality measurement data. Consequently, unsupervised approaches that do not need such reference data have gained increased attention recently   \cite{hendriksen2020noise2inverse, Lehtinen2018, Kim2021}. Most notably, and with relevance to this study, are denoising approaches that can be trained with noisy reference data of either paired  \cite{hendriksen2020noise2inverse, Lehtinen2018} or unpaired images \cite{Kim2021}. If paired images are used, both instantiations must have uncorrelated noise. In tomographic reconstruction, this has been realized as Noise2Inverse \cite{hendriksen2020noise2inverse}, where by angular binning non-overlapping subsets of the measured sinogram are used to produce reconstructions of the same target with uncorrelated noise. Nevertheless, this approach is expected to yield a reconstruction equivalent to the filtered backprojection from noise free data and hence is limitedly applicable to the sparse angle scenario\rev{, as it retains streaking artefacts as well.} 
To overcome this problem, we adopt a different strategy tailored to our application of sparse multi-spectral CT. First, instead of taking subsets of the sinogram, we produce two reconstructions from the full measurement with different filter functions in the filtered backprojection (FBP), which are used as training pairs. Second, we add a total nuclear variation loss to the training to couple spectral channels and suppress remaining streaking artefacts.

\rev{Let us shortly discuss other unsupervised methods that could be adapted with some modifications. Generative adversarial networks still need (unpaired) reference data to train the discriminator \cite{yang2018low,ma2020low}, which are not readily available in our case. SURE (Stein's unbiased risk estimate) based approaches make use of the noise statistics in the image \cite{metzler2018unsupervised,nguyen2020hyperspectral,kim2020unsupervised}. Here, we have strongly varying noise between spectral channels and even within the same image (air and metallic) and hence necessitates careful adjustment of the method. Lastly, the Deep Image Prior (DIP) \cite{ulyanov2018deep} could be used with a suitable regularizer to suppress the streaking artefacts \cite{baguer2020computed}, but is known to be computationally demanding and would need an involved pretraining procedure for 3D spectral CT \cite{barbano2021deep}. }

\rev{This paper is structured as follows.} The proposed unsupervised denoiser for flexible sparse spectral multi-bin PCD-CT will be discussed in Section \ref{sec:unsupervised}. We will also compare the proposed method to iterative total nuclear variation (TNV) reconstructions \cite{rigie2015joint} and a supervised denoiser using a reference reconstruction, following \cite{Jin2017,mustafa2020sparse}.
We will apply the learned methods on synthetic 2D multi-channel (64 energy channels) sparse PCD projection data, described in Section \ref{sec:dataSection}. Finally, we utilize the openly available MUSIC experimental dataset \cite{Kehl2018} to show how the developed method performs for real experimental data.
We present results, assess the computation time and reconstruction quality in Section \ref{sec:results}. We then discuss the proposed method in Section \ref{sec:discussion} and present final conclusions in Section \ref{sec:conclusions}.

\section{Unsupervised training for sparse multispectral tomographic data}

There is recent rising interest in unsupervised learned reconstruction methods, as in many applications there is only limited ground-truth information available, which either needs to be simulated or computed from high accuracy scans. 
Obtaining reference scans can be especially difficult for multi-spectral CT, where strong noise occurs in very low and high energy channels due to lower photon counts, and emphasized streaking artefacts appear for strongly attenuating objects in the low energy channels. Additionally, depending on the photon energies we observe different noise statistics, which makes fine-tuning of reference reconstructions a tedious and difficult task.

Consequently, we investigate in the following the suitability of a class of learned denoisers that can operate without ground-truth data and can learn from the same dataset by exploiting uncorrelated noise. As various studies have recently reported, a denoising network can be trained with pairs of images with different noise instantiations \cite{hendriksen2020noise2inverse, hendriksen2021deep, Lehtinen2018}.  \rev{The observation by \cite{Lehtinen2018}, introduced as \emph{Noise2Noise}, is that when given two realisations of a noisy image $x^{\delta_1}=x+\delta_1$ and $x^{\delta_2}=x+\delta_2$ with independent draws of $\delta_i\sim\pi_{\text{noise}}$ one can  reformulate the supervised training setting using the two noise corrupted images. 
For zero-mean noise $\mathbb{E}[\noise]=0$ the expected prediction error 
\begin{equation}\label{eqn:noise2noise_pred} 
    \theta^* = \argmin_\theta \mathbb{E}_{x,\delta_1,\delta_2}\left[ \| \Lambda_\theta(x+\delta_1)  - (x+\delta_2) \|^2_2 \right]
\end{equation}
for a class of regression networks $\Lambda_\theta$ with parameters $\theta$ is then minimized by the same (regression) network $\Lambda_{\theta^*}$ as in the supervised case.}

\rev{Based on the same observation,} the \emph{Noise2Inverse} \cite{hendriksen2020noise2inverse,hendriksen2021deep} method has been introduced for single energy CT setting using two, or more, reconstructions with uncorrelated noise.
Such reconstructions can be easily produced by angular binning of the projection data into non-overlapping subsets and by reconstruction with FBP. This has been shown to work well for dense angular sampling, when no streaking artefacts are present. 
Nevertheless, for sparse imaging scenarios, streaking artefacts will remain. Additionally, varying noise statistics in the multi-spectral case can be problematic for the approach.
Thus, we need to take a different angle on the unsupervised training of a denoiser that also works in the case of sparse multi-spectral data. Here, the trained denoiser also needs to act as an artefact removal network. In the following, we will extend the unsupervised denoising approach \rev{based on the Noise2Inverse} by producing two reconstructions from the same set of projections using different filters in the backprojection, instead of angular binning. 

In the following we will denote the ground-truth tomographic image by $x$ and the forward model describing the projection geometry is given by $A$, the precise imaging environment will be described in Section \ref{sec:MUSIC_data}. Then the measurement data $y$ is obtained as
\begin{equation}\label{eq:forwardMod}
    Ax + \delta=y,
\end{equation}
where $\delta$ denotes noise in the measurement process. A reconstruction $x_\text{FBP}$ can be obtained from $y$ by filtered backprojection (FBP), where visible features in the obtained reconstructions will depend on the type of filtering employed. That means, we can decide on whether to emphasize low or high frequencies, influencing the image characteristics, as will be discussed below in Section \ref{sec:unsupervised}.

The aim in the following is to to train a network $\Network_\theta$, with parameters $\theta$, that can improve the reconstruction quality of the obtained filtered backprojection, such that
\begin{equation}\label{eq:networkMap}
 x_{\text{rec}}= \Network_\theta(x_\text{FBP}).   
\end{equation}
The primary question is how training of the denoising network can be performed, especially considering one has no or little ground-truth data available. 

\subsection{Supervised training}\label{sec:supervised}

Let us first discuss the supervised approach \rev{when a known ground-truth or high-quality reconstruction is available  \cite{Kang2017,Jin2017}, as it is the most established way to train the network in \eqref{eq:networkMap} and it will also serve as the basis for our unsupervised setting}. \rev{Here, a reference} high-quality reconstruction should be ideally given from dense angular sampling under high dose. If no such data is available, one would need to create a reference reconstruction that is considered the gold-standard for the given scenario. The network is then trained to reproduce a similar result but considerably faster than the reference method.
For the application of spectral CT, such a reference reconstruction could be given by an iterative reconstruction with a total nuclear variation penalty term \cite{rigie2015joint}, 
which introduces a gradient-coupling effect between energy channels by promoting rank-sparsity of the coupled derivatives.

If given such training pairs of noisy reconstruction and a reference for training $\{x_\text{FBP}^{(n,b)}, x_\text{ref}^{(n,b)}\}$, where $n\in[1,N]$ denotes a sample from the training pairs and $b$ the spectral channels. One can then train the \rev{supervised denoising network by minimizing a suitable loss function.} In this study we will use the squared $\ell^2$-norm
\begin{equation}\label{eqn:l2loss_super}
L(x_\text{FBP}^{(n,b)},x_\text{ref}^{(n,b)}) = \bigl\|\Network_\theta(x_\text{FBP}^{(n,b)})  - x_\text{ref}^{(n,b)} \bigr\|_2^2.
\end{equation}
The primary advantage of the supervised denoiser setup lies in its simplicity and its computational speed in the reconstruction phase, as it only requires a single reconstruction by filtered backprojection for each spectral channel and a computationally efficient evaluation of the network, after the training has been performed. On the downside, the performance of the network is governed by the quality of the reference reconstructions in the training data. As such, unsupervised approaches are desirable that can extract relevant information from only the noisy data \rev{and are not limited by the quality of the available reference}. In our experiments, the FBP reconstruction for the supervised training had  Hann filtering with a frequency scaling of 0.6.

\subsection{Unsupervised training}\label{sec:unsupervised}

We now aim to train a denoising and artefact removal network with only noisy pairs \rev{following the Noise2Inverse approach. Such a} denoising network can be efficiently trained when two noisy instances of the same tomographic image are available.
Here, we propose a strategy to produce two reconstructions from the same sinogram. To achieve this we will utilize different filter functions in the filtered backprojection.

We first realize, that given a noisy reconstruction $\widetilde{x}=x+\delta$, we ideally would like to have access to the noise component $\delta$. \rev{Then for training one could separate it into two components $\delta_1$ and $\delta_2$, such as} low and high frequency components by separation in Fourier space using high-pass and low-pass filtering. Similarly, a separation of frequencies could be achieved by employing different filters in the filtered backprojection. \rev{Specifically,} a reconstruction can be expressed by 
\begin{equation}\label{eq:FBP}
x_\text{FBP}=A^\ast(h \ast y ) = A^\ast(h \ast (Ax + \delta) )
\end{equation}
where the obtained data $y$ is filtered by a convolutional filter before backprojected with the corresponding adjoint operator $A^\ast$, also called the backprojection,  for the given measurement geometry (including measurement noise).
In learning methods for low-count data $h$ is often chosen to be a low-pass filter, such that primarily high-frequency noise is removed from the data before backprojection is performed. Our aim is to utilize two different filters to produce two reconstructions, where one has an emphasis on low-frequency features and the other leaves high-frequency features intact.

\begin{figure}[t!]\centering
\centering
\includegraphics[width=0.5 \textwidth]{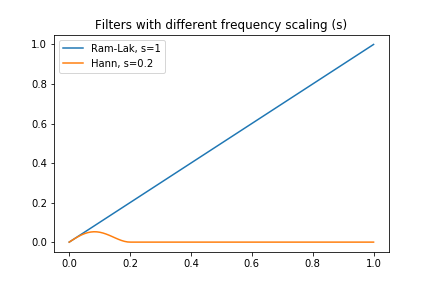}
\caption{Used filter functions in frequency space to create the two training pairs for the Low2High method.}
\label{fig:filterFunc}
\end{figure}


Unfortunately, we can not separate the data into high and low frequencies as this would result in a loss of structural similarity of image features in the reconstruction. But, rather we can obtain one smoothed (almost) noise and streaking artefact free reconstruction from the low-pass filter and one reconstruction including the high frequency noise and sharp image features. The resulting filters used in this study are shown in Fig. \ref{fig:filterFunc} with their normalized representation in frequency space. For this study we utilize a Ram-Lak filter for the high frequency features, given by the identity with frequency scaling $s=1$, 
\[
h_\text{Hi}(k)=k, \text{ for } k\in[0,1],
\]
which, in fact, means we do not cut-off any noise components before we perform the filtered backprojection. For the low-pass reconstruction, we use a Hann filter with a frequency scaling $s$, that is
\[
h_\text{Lo}(k)=k \left(\cos\left(\frac{k \pi}{ 2  s}\right)\right)^2, \text{ for } k\in[0,s]
\]
and $h_\text{Lo}(k)=0$ otherwise. 
In this study, frequency scaling of $s=0.2$ was applied for all experiments.
We note that also other choices of frequency scaling $s$ and the filter functions can be chosen, if needed. 
For the unsupervised training task we then obtain two reconstructions $x_\text{Lo}$ and $x_\text{Hi}$ that serve as training pairs. We can then employ the same loss function in Eq. \eqref{eqn:l2loss_super} with the corresponding training pairs $\{x_\text{Lo}^{(n,b)},x_\text{Hi}^{(n,b)}\}$. \rev{We note, that training over multiple samples or spectral channels is necessary for the varying noise realizations.}
We will call our approach \emph{Low2High}. An illustrative result of this approach is shown in Fig. \ref{fig:syntTest}. As it can be observed for dense angular sampling the network nicely sharpens the low-pass filtered reconstruction without reproducing the noise. 
Whereas in the sparse case, clear streaking artefacts are reintroduced, which also result in a residual noise pattern in the image. Even though the low-pass filtered version suppresses streaking artefacts, these reappear partly during the training procedure of the denoiser. \rev{This can be explained by the prediction error in Eq. \eqref{eqn:noise2noise_pred} and zero-mean condition on the noise, which is clearly not satisfied for streaking artefacts.}
Thus, in the following we will address this issue with an additional penalty term for the unsupervised training task.

{\newcommand{\showpic}[3]{%
\begin{tikzpicture}
\draw (0,3.2) node [anchor=south] {\phantom{f}#1\phantom{g}};%
\draw (0,2.8) node [anchor=south] {\phantom{f}#2\phantom{g}};%
\draw (0,3) node [anchor=north] {\includegraphics[width=2.8cm]{synTest/#3_512_40}};%
\draw (0,0) node [anchor=north] {\includegraphics[width=2.8cm]{synTest/#3_32_40}};%

\end{tikzpicture}\hspace*{-2mm}%
}

\begin{figure*}[tb!]
\centering
\showpic{Phantom}{}{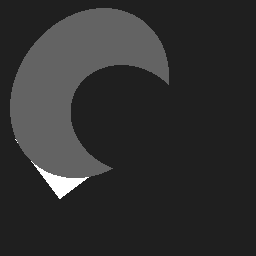}%
\showpic{Input}{Hann filter}{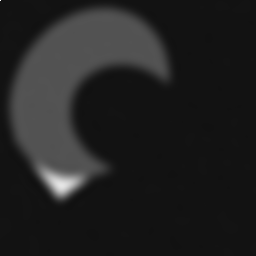}%
\showpic{Training pair}{Ram-Lak filter}{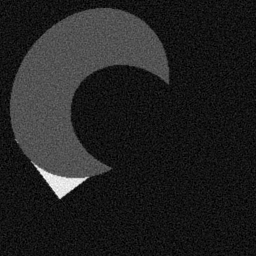}%
\showpic{Low2High}{}{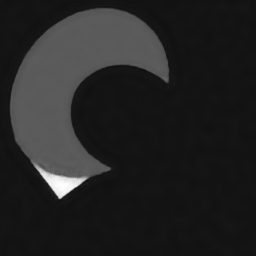}%
\showpic{Low2High}{TNV: $\lambda=0.005$}{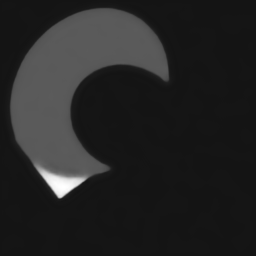}%
\showpic{Low2High}{TNV: $\lambda=0.02$}{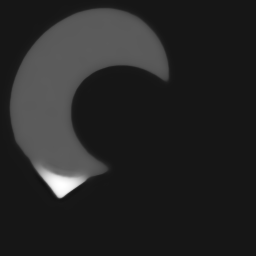}%
\put(-495,20){\rotatebox{90}{32 angles}}
\put(-495,110){\rotatebox{90}{512 angles}}
\caption{\label{fig:syntTest} Illustration of the Low2High training strategy. Training pairs of low frequency by a Hann filter ($s=0.2$) and high frequency with a Ram-Lak filter ($s=1$) are shown in the 2nd and 3rd column, respectively. Result by training only with the two pairs as Low2High are shown in the 4th column. The 5th and 6th column show denoising results with added TNV penalty and different weighting factors.}
\end{figure*}}

\subsubsection{Spectral channel coupling with total nuclear variation}

In the case of sparse spectral computed tomography we can exploit that edges between spectral channels will be aligned. Thus, to avoid that the Low2High training procedure reintroduces streaking artefacts, while keeping edges of the target, we add a total nuclear variation penalty to the training procedure. The use of a total nuclear variation (TNV) penalty is a highly successful technique to couple the gradients between spectral channels by requiring rank sparsity of the Jacobian \cite{rigie2015joint}. Similar penalty terms based on the coupling of edges have been also successfully used in multi-modality imaging \cite{ehrhardt2014joint, knoll2016joint}. Thus, in the following we will employ a TNV penalty in our training task to improve the results of the unsupervised training procedure.

Given the same paired training data $\{x_\text{Lo}^{(n,b)},x_\text{Hi}^{(n,b)}\}$ as for the unsupervised training task above, we now add a total nuclear variation penalty to the loss function in Eq. \eqref{eqn:l2loss_super}, that is
\begin{equation}\label{eqn:loss_unsuper}
\begin{split}
L(x_\text{Lo}^{(n,b)},x_\text{Hi}^{(n,b)}) = \bigl\|\Network_\theta(x_\text{Lo}^{(n,b)}) & - x_\text{Hi}^{(n,b)} \bigr\|_2^2 \\
&+ \lambda \bigl\|J(\Network_\theta(x_\text{Lo}^{(n,b)}))\bigr\|_{1,\star},
\end{split}
\end{equation}
where $\lambda$ is a weighting term for the TNV penalty. The TNV penalty can be efficiently computed as the $\ell^1$-norm of the singular values of the Jacobian $J$ of the output from the network. Most importantly, this can be easily added to the training procedure as an additional loss function, where automatic differentiation can be performed to update the parameters $\theta$.

Finally, we note that the low-pass filter herein has been chosen as a starting point such that streaking artefacts are suppressed in the input to the network, the employed TNV penalty in the training procedure then successfully prevents these to reappear. Nevertheless, we can see in Fig. \ref{fig:syntTest} that for lower values of the weighting parameter $\lambda$ we still see some residual artefacts in the background. For higher values we get a clean image, but edges between different regions start to be smoothed out. For the rest of this study, we will chose a value of $\lambda=0.005$ to balance this behaviour, for an emphasis on retaining edges of the target.


\section{Synthetic and experimental data}\label{sec:dataSection}

Two datasets were adopted for this study: a measured, openly available, Multi-Spectral Imaging via Computed Tomography (MUSIC) dataset and synthetic images dataset using the TomoPhantom python software package \cite{Kehl2018, Kazantsev2018a}.

\subsection{Open source MUSIC dataset}\label{sec:MUSIC_data}
Detailed description of the openly available MUSIC dataset can be found from Kehl et al. \cite{Kehl2018} and the dataset can be downloaded from\footnote{\url{http://easi-cil.compute.dtu.dk/index.php/datasets/music/}}. Here, we briefly summarize the MUSIC dataset. The dataset was designed for security scanning assessment, and it consists of seven sample tubes containing different items from fluids, fruits, threat, and non-threat item categories (Table \ref{tab:table_MUSIC_splits}). Samples were imaged with a 2-D table-top  spectral CT device. The device had a MultiX ME-100 V2 cadmium telluride (CdTe) X-ray detector, which has 128 energy bins and an X-ray microfocus source with peak kilovoltage set to 160 kVp. The PCD consists of two modules of 128 pixels yielding a total of 256 pixels with 0.77 mm  pixel size. The scan geometry was set such that the source-to-detector and source-to-object distances were 115.55 mm and  57.50 mm, respectively. Thirty-seven projections were collected during acquisition \cite{Kehl2018}.

\begin{table}[b]
\caption{Summary of MUSIC dataset splits; Number of slices (\textit{N}) and samples in the sets.}
\begin{center}
\begin{tabular*}{80mm}{ l c c }
  \hline
  Set & \textit{N} & Sample name  \\ 
  \hline
 Train & 929 & Fluids, Fruits, Sample$\_$23012018,\\
  & & Sample$\_$31102016\\
 Validation & 382 & NonThreat, Threat\\
 Test & 278 & Sample$\_$24012018 \\
 \hline
\end{tabular*}
\label{tab:table_MUSIC_splits}
\end{center}
\end{table}

\begin{figure*}[!b]\centering
\centering
\includegraphics[width=175mm]{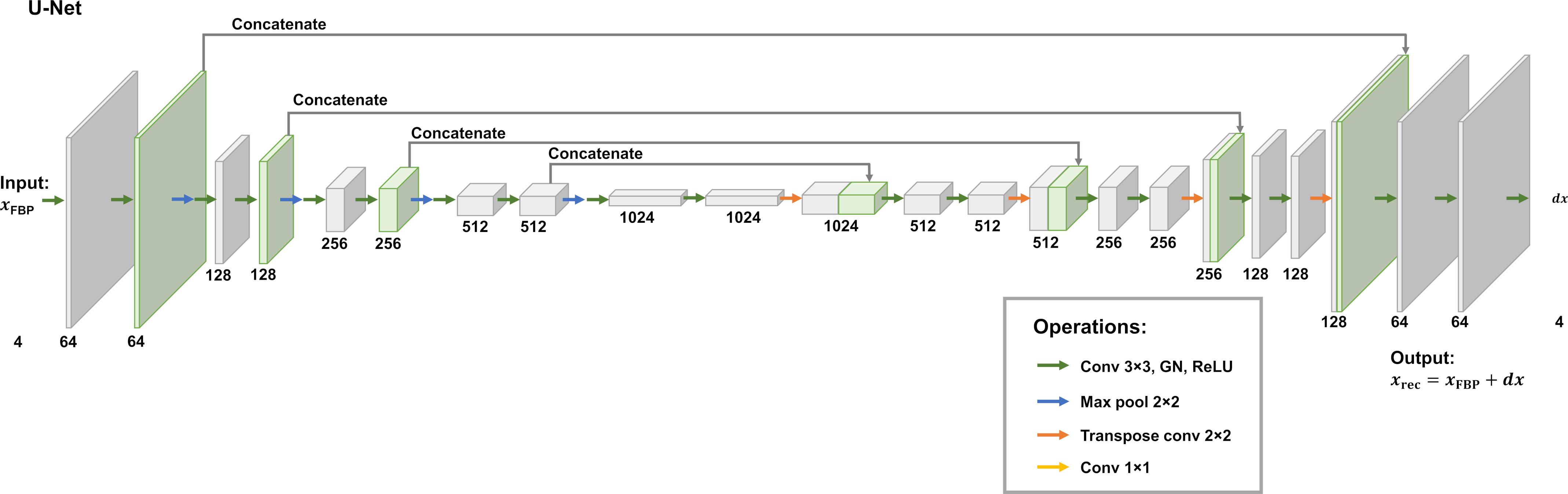}
\caption{Diagram of the residual UNet architecture used as post-processing network. The inputs reconstructions are given by a filtered backprojection (FBP) and the network computes a residual update as additive correction term to the initial reconstruction. We used group normalization (GN) here in the convolutional layers.}
\label{fig:networks}
\end{figure*}

After the projection data was downloaded, it was processed as follows. The number of spectral channels was reduced from 128 energy bins to 64 channels from 55 keV to 124 keV, and projection data from each sample was manually cropped such that the empty tube ends containing no material were excluded.

For the application of spectral computerized tomography, we compute a computationally expensive iterative joint (spectrally coupled) reconstruction using total nuclear variation regularization (IR TNV) which introduces a gradient-coupling effect for energy channels as it encourages rank-sparsity. IR TNV is used as the reference ground truth ($x_\text{ref}$) for sparse MUSIC dataset.  
The reconstruction problem is then given as by
\begin{equation}\label{eqn:IR+TNV}
x_{\rm{ref}}=\argmin_x  \frac{1}{2}\bigl\|Ax - y\bigr\|_2^2 + \alpha \bigl\|(Jx)\bigr\|_{1,\star},
\end{equation}
where $y$ is the logarithm transformed flat-field corrected, i.e., linearized projection data (sinogram) and $A$ denotes the projection matrix, $Jx$ is the Jacobian, and $\bigl\|(Jx)\bigr\|_{1,\star}$ is the total nuclear variation penalty, the parameter $\alpha>0$ controls the regularization strength. Reconstructions were performed in 2D slices such that four adjacent energy slices were fed into the algorithm yielding 16 batches for all 64 energy channels.

Pre-processing and reconstructions were conducted using python and the operator discretization library (ODL) library \cite{Adler2017}. ODL library's  Douglas-Rachford primal-dual solver was applied for this problem with step sizes $\tau$ = 0.01 for indicator box function and $\sigma_1$ = 0.01 for data fidelity and  $\sigma_2$ = 0.5 for TNV. The regularization parameter $\alpha$ was fixed to 100 for each energy channel and the solver was run for 50 iterations using FBP as initial solution.

After computing the reference reconstructions the training, validation, and test set splitting was performed for the dataset (Table \ref{tab:table_MUSIC_splits}), where samples tubes were not split between sets. We note that each sample consists of 64 channels to make up the whole dataset size.

\subsection{Synthetic multi-energy dataset}
A simulated synthetic dataset was created with the TomoPhantom python package \cite{Kazantsev2018a}. The simulation geometry was kept similar to the MUSIC data acquisition.  The 2D synthetic dataset consisted of random sized, positioned and rotated ellipsoidal or rectangular objects. The number of structures varied from 0 to 5 per slice and the objects were enforced to be contained in the circular reconstruction domain. 
 If two or more ellipses overlapped, the most recently generated ellipse was kept in the overlapping area. Finally, the objects extending outside the field-of-view of a 256-pixel detector, were cropped from the image.
 
A simple effective linear attenuation (LAC) model was applied to generate spectral information
\begin{equation}\label{eqn:LAC_model}
\mathrm{LAC} = A e^{-BE},
\end{equation}
where $A \sim U(0, 0.03)$, $B\sim U(0, 10)$, $E$ are the 64 energy channels from 55 keV to 124 keV. The ranges for uniform distributions were based on the MUSIC dataset as the maximum effective linear attenuation of 0.03 was estimated from aluminium bar region present in the reconstructed dataset. 

After the spectral information was added to the synthetic dataset, we simulated 37 angular projections using ODL library with ASTRA backend extension \cite{VanAarle2016, Aarle2015} with the MUSIC geometry. The number of projections was matched to the MUSIC dataset. 2$\%$ random Gaussian noise was added to the simulated projections. For training, validation and testing 1200, 100 and 100 samples were realized, respectively.

\subsection{Network implementation}

For our implementation we used an adaption of the popular UNet architecture \cite{ronneberger2015u} with a residual update similarly to \cite{Jin2017}, as illustrated in Fig. \ref{fig:networks}. Implementation differed from  \cite{Jin2017} such that  group normalization  (number of groups $=$ 1) was used instead of batch normalization. In particular, we trained a 2-D UNet which operates on 1 and 4 adjacent energy channel input images for supervised and unsupervised learning, respectively. The energy channels were randomly selected from the spectral channels of a slice during training. For synthetic dataset the input and reference data was normalized prior feeding to network but for MUSIC dataset the input and reference was not normalized. 

For supervised and unsupervised training of the UNet we applied approaches explained in the  \ref{sec:supervised} and \ref{sec:unsupervised}, respectively. We note, that $x_{\rm{ref}}^{(n,b)}$ here is either the ground-truth synthetic dataset or the reference reconstruction (IR TNV) for the MUSIC dataset obtained by \eqref{eqn:IR+TNV}. We note, that in the latter, the $x_{\rm{ref}}^{(n,b)}$ might include a bias due to the chosen reconstruction algorithm to obtain a gold-standard reference. Here the randomization over spectral channels will be essential to counteract possible overfitting issues to the reference method as will be discussed later.

Networks were implemented under the PyTorch deep learning framework (v.1.2.0). For all networks and both datasets the training was performed with batch size of 1, using a default Adam optimizer for 50 epochs, with learning rate of $10^{-4}$.  A cosine annealing  scheduler with default parameters was applied after each epoch to reduce the initial learning rate. The supervised UNet denoiser model was trained on training set and the unsupervised Low2High method was trained using test set only.

The neural networks were trained on a  CSC IT center for science supercomputer with  
NVIDIA Volta V100 GPU (24 GBit memory) and 10 CPU threads. Training of UNet  and UNet with TNV penalization network for the MUSIC dataset took 34 minutes and 2 hours and 4 minutes, respectively. The TNV had to be computed on CPU, as at the time of this study no GPU support for computing the singular value decomposition was available in PyTorch. After the neural network models were trained, the reconstruction were fast to compute compared to IR TNV (Table \ref{tab:table_Ctimes}).

\begin{table}[h!]\centering
\caption{Computational time of ten 64 spectral channel slice images on MUSIC dataset.}
\label{tab:table_Ctimes}
\begin{tabular}{ l c } 
\hline
Method & Computation time $($HH:MM:SS$)$ \\
\hline
IR TNV    & 0:32:41 \\
UNet      & 0:01:12 \\
Low2High  & 0:01:09 \\
\hline
\end{tabular}
\end{table}

\subsection{Image quality assessment}
\begin{figure}[h!]\centering
\centering
\includegraphics[width=45mm]{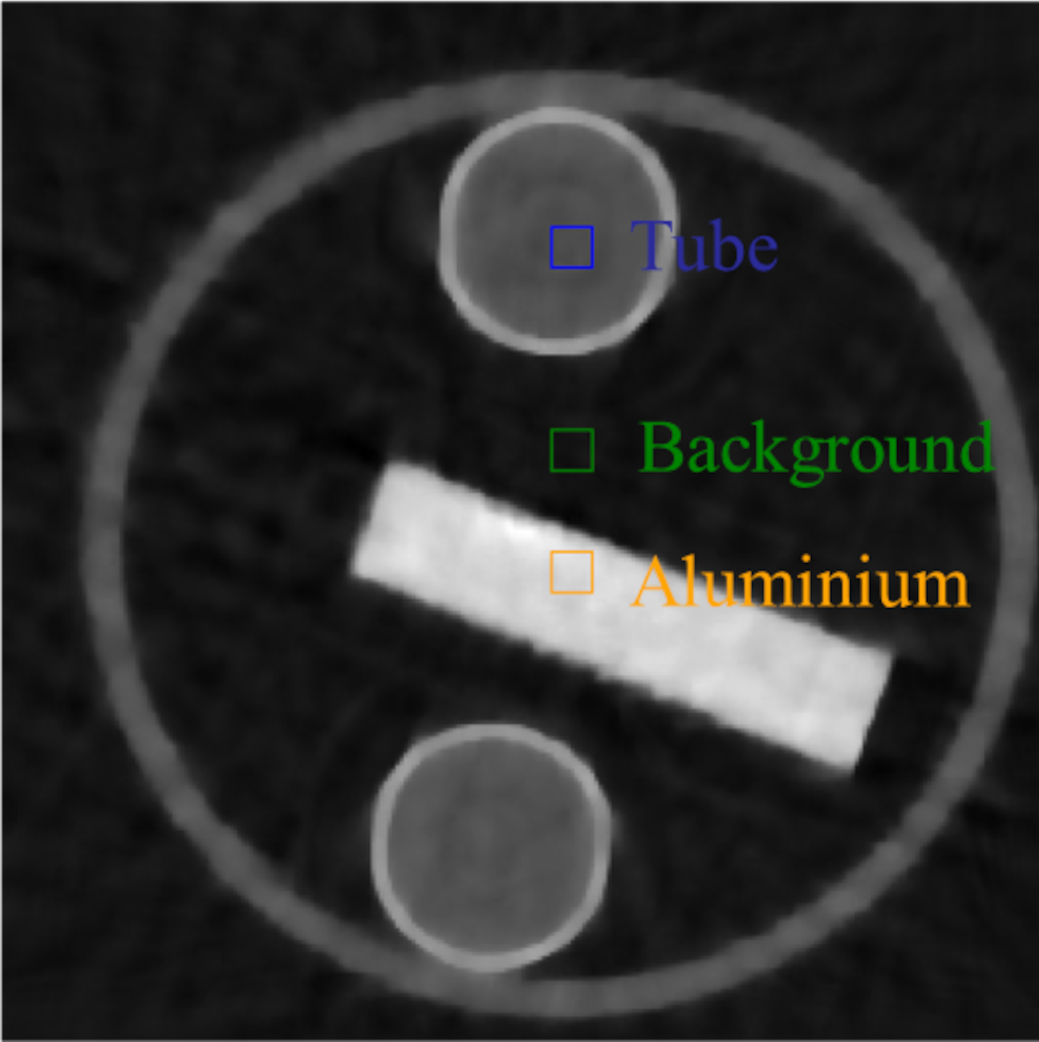}
\caption{\rev{Region of interest locations chosen for CNR and SNR computations of tube (low contrast) and aluminium (high contrast) regions. The illustration slice image presented here is the IR TNV average over all energy channels.}}
\label{fig:ROI-Locations}
\end{figure}

Structural similarity index (SSIM) and peak signal-to-noise ratio (PSNR) were determined from the reconstructed test set images. In addition,  channel-wise image quality was evaluated. For the synthetic dataset  the ground truth images were available. For MUSIC dataset iterative total nuclear variation (IR TNV) was chosen as the reference reconstruction. To qualitatively assess the preservation of the effective linear attenuation coefficients (LACs), we determined LACs of different materials from example slices for MUSIC dataset.
\rev{In addition, one representative slice image was chosen from the MUSIC test dataset containing both low (tube) and high (aluminium) contrast regions (Fig. \ref{fig:ROI-Locations}). From those regions, contrast-to-noise ratio (CNR), and signal-to-noise ratio (SNR) were computed.}

\begin{figure*}[t!]
\centering
\includegraphics[width=181mm]{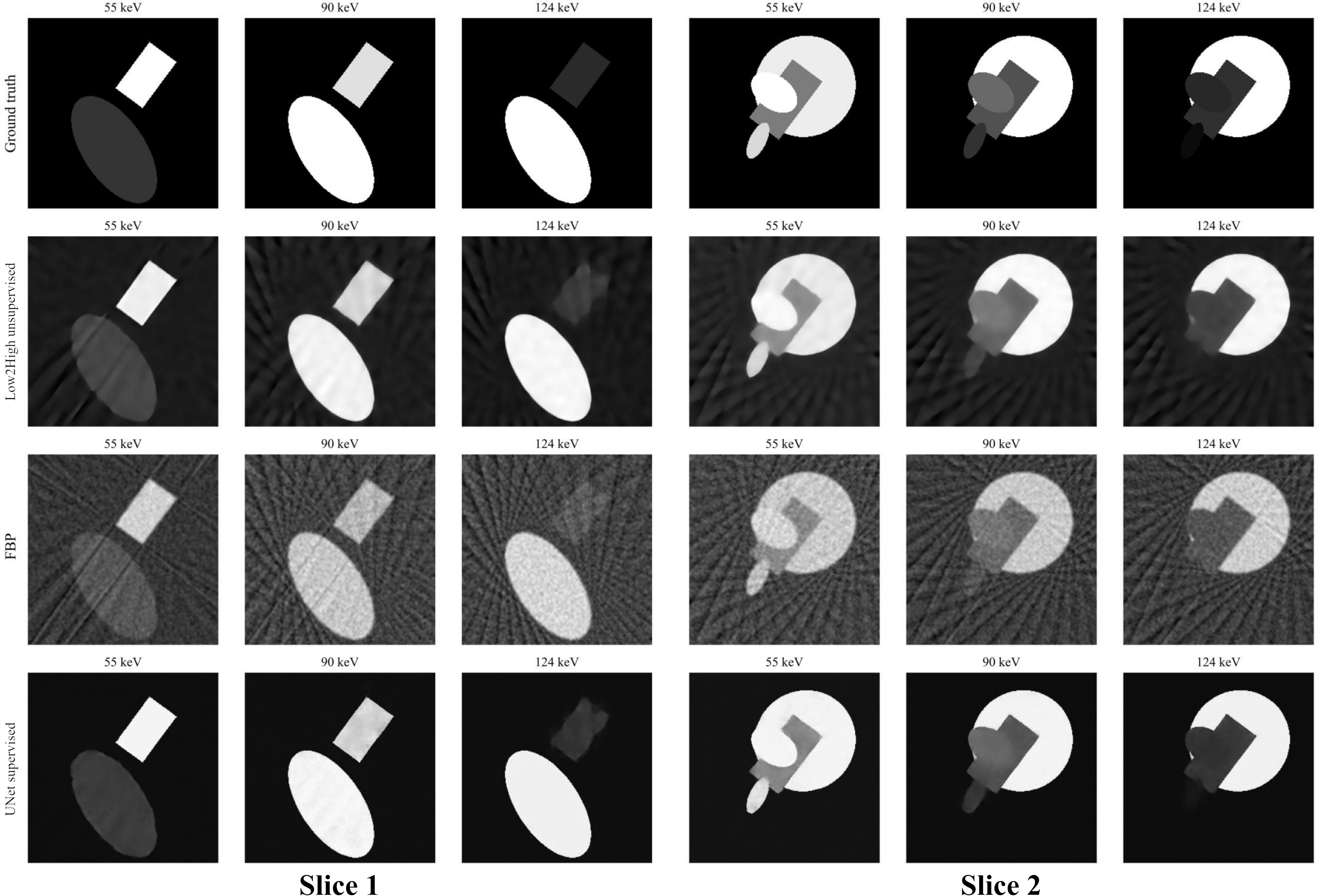}
\caption{Channel-wise reconstructions for sparse synthetic dataset of two randomly selected slices from test set. The unsupervised Low2High and supervised UNet methods were able to suppress the streaking artefacts. In the unsupervised approaches the streaking artefacts were not so effectively suppressed at low energies, however in the noisy high energy channel 124 keV results are improved when compared to FBP.}
\label{fig:Chan_recons_Synthetic}
\end{figure*}

Finally, maximum intensity projection (MIP) images were computed from a sum of reconstructed energy channel  images for the test set of the MUSIC dataset.

\begin{figure*}[t!]\centering
\centering
\includegraphics[width=181mm]{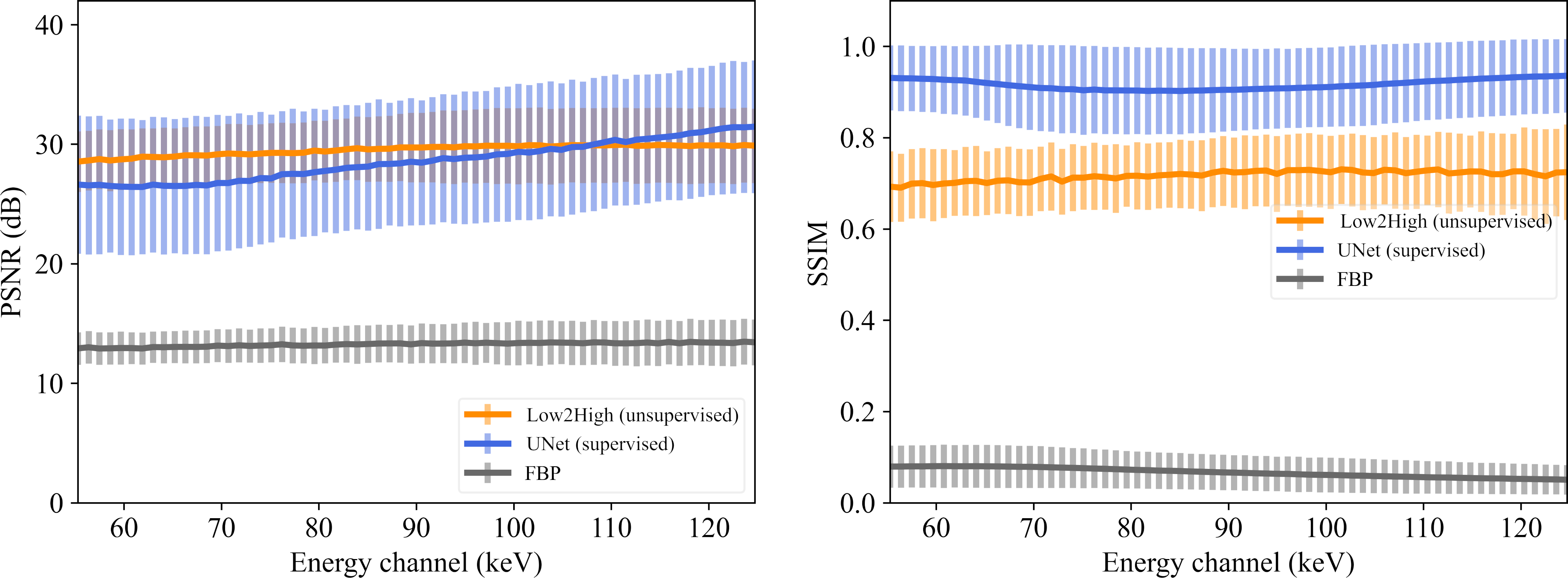}
\caption{Channel-wise PSNR and SSIM (mean $\pm$ SD) for spectral synthetic dataset. The highest PNSR and SSIM was observed with UNet.}
\label{fig:Chan_PSNR_Synthetic}
\end{figure*}

\section{Results}\label{sec:results}

\subsection{Image quality assessment}
\subsubsection{Synthetic spectral dataset}
The supervised UNet denoising model had the best SSIM when compared with FBP and unsupervised Low2High method (Table \ref{tab:table_IQ_SYN}). The Low2High method however, was able to suppress streaking artefacts of FBP reconstructions and clearly surpassed the FBP reconstruction in image quality (Fig. \ref{fig:Chan_recons_Synthetic}). Interestingly, in the PSNR assessment both supervised and unsupervised methods had similar performance (Table \ref{tab:table_IQ_SYN} and Fig. \ref{fig:Chan_PSNR_Synthetic}).

In our experiments, we found that model input normalization for synthetic dataset improved the overall image quality of different energy channels and the technical image quality was preserved for different reconstruction methods energy channel-wise (Fig. \ref{fig:Chan_PSNR_Synthetic}).

\begin{table}[h]\centering
\caption{Structural similarity index (SSIM), and peak-signal-to-noise (PSNR) for different reconstruction methods for Spectral synthetic dataset (mean $\pm$ SD).}
\label{tab:table_IQ_SYN}
\csvreader[ 
tabular={lcc},
    table head=\hline Method & SSIM & PSNR (dB) \\\hline,
    late after last line = \\\hline
     ]
{IQ_res_SYN_pm.csv}{name = \name, SSIM=\SSIM, PSNR=\PSNR}
{ \name & \SSIM & \PSNR}
\end{table}

\begin{figure*}[b]\centering
\centering
\includegraphics[width=181mm]{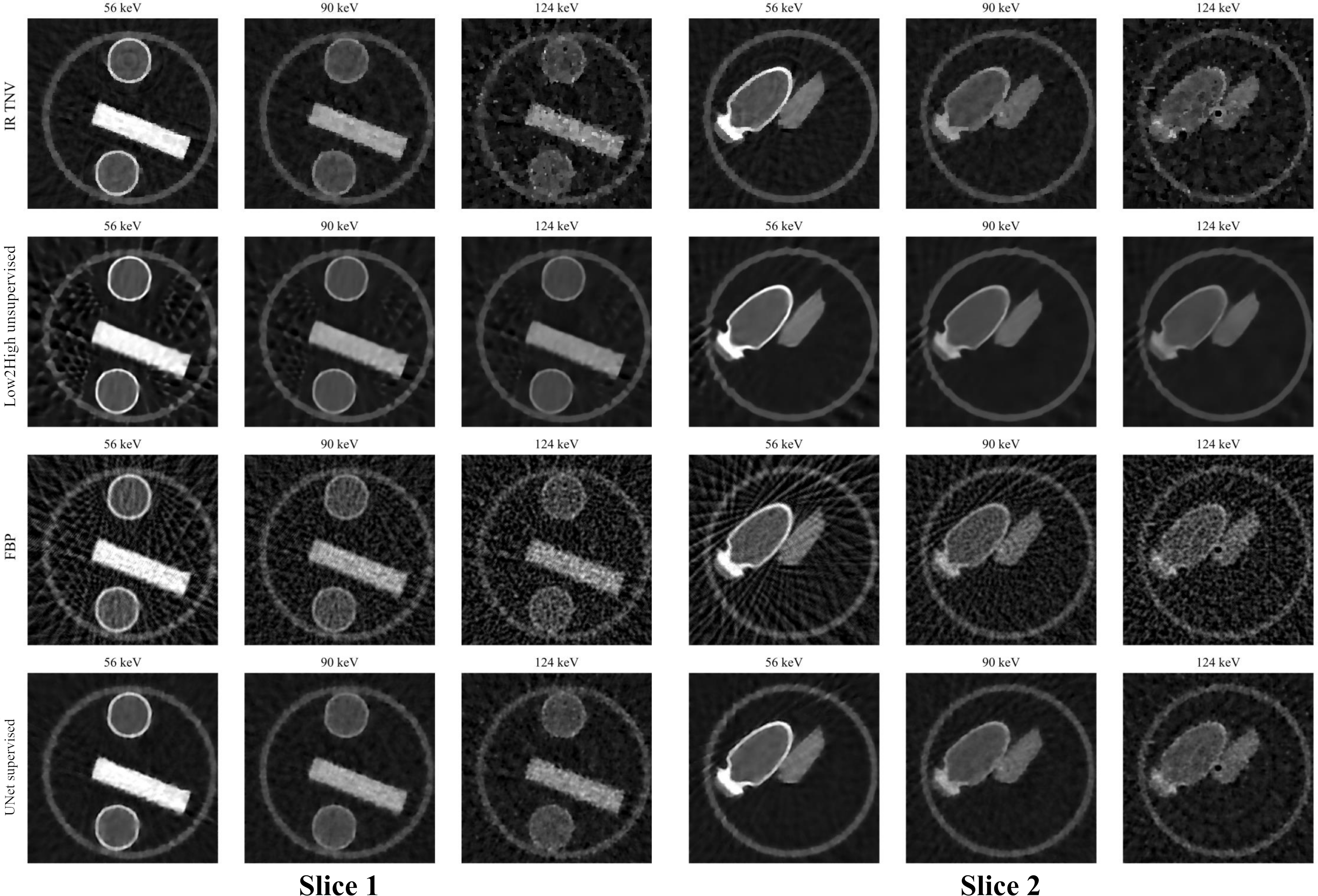}
\caption{Channel-wise reconstructions for MUSIC dataset of two slices. The unsupervised Low2High method for test set was able to suppress streaking artefacts most effectively.
IR TNV, supervised and unsupervised  methods were able to suppress the streaking artefacts arising from aluminium bar low energy channel 56 keV as shown in Slice 1. Windowing is set to [-0.002, 0.03].}
\label{fig:Chan_recons_MUSIC}
\end{figure*}

\subsubsection{Spectral MUSIC dataset}
After training, the learning-based methods produced qualitatively similar reconstruction quality as the reference IR TNV method as shown in (Fig. \ref{fig:Chan_recons_MUSIC}). 
Interestingly, at the noisier high  energy channels, the learning based methods were able to effectively suppress noise more than the reference IR TNV method when evaluated qualitatively (Fig. \ref{fig:Chan_recons_MUSIC}), even for the supervised approach that was trained on the reference IR TNV reconstructions. This can be interpreted as a regularizing effect of training with all spectral channels and effectively reduces the bias introduced by the reference reconstructions. The unsupervised reconstructions Low2High showed excellent noise suppression exceeding the supervised method in image quality.
 However, the FBP related streaking artefacts were not perfectly suppressed in the low energy channels of the unsupervised Low2High method (Fig. \ref{fig:Chan_recons_MUSIC}).

 \begin{table}[t]\centering
\caption{Structural similarity index (SSIM), and peak-signal-to-noise (PSNR) for different reconstruction methods for MUSIC dataset (mean $\pm$ SD). The ground truth reference image was the IR TNV.}
\label{tab:table_IQ_MUSIC}
\begin{filecontents*}{IQ_res_MUSIC_pm.csv}
,name,SSIM,PSNR
0,UNet (supervised),0.643 $\pm$ 0.048,30.2 $\pm$ 1.5
1,Low2High (unsupervised),0.454 $\pm$ 0.080,27.6 $\pm$ 1.8
2,FBP,0.315 $\pm$ 0.044,25.2 $\pm$ 1.4

\end{filecontents*}
\csvreader[
tabular=lcc,
    table head=\hline Method & SSIM & PSNR (dB) \\\hline,
    late after last line = \\\hline
    ]
{IQ_res_MUSIC_pm.csv}{name = \name, SSIM=\SSIM, PSNR=\PSNR}
{\name & \SSIM & \PSNR}
\end{table}

\begin{table}[b!]
\caption{\rev{Qualitative image measures presented as mean values over energy channels on one representative slice image from MUSIC dataset. Shown are: contrast-to-noise ratio (CNR), and signal-to-noise ratio (SNR).}}
\begin{center}
\resizebox{\columnwidth}{!}{
\begin{tabular*}{75mm}{ l c c c c }
  \hline
  Method & CNR$_{\rm{T}}$ & SNR$_{\rm{T}}$  & CNR$_{\rm{Al}}$ & SNR$_{\rm{Al}}$ \\ 
  \hline
 Low2High & \bf{31.3}  & \bf{32.2} & 84.0       & 84.9 \\
 FBP      &  5.0       & 5.2       & 14.0       & 14.2 \\
 UNet     &  30.7      & 31.2      & \bf{91.2}  & \bf{91.7} \\
 IR TNV   &  20.6      & 20.8      & 61.3       & 61.5 \\
 \hline
 \multicolumn{5}{l}{\footnotesize{T: tube (low contrast target), Al: 
Aluminium (high contrast target)}}
\end{tabular*}}
\label{tab:table_MUSIC_qualitative_IQ}
\end{center}
\end{table}

\begin{figure*}[t!]\centering
\includegraphics[width=181mm]{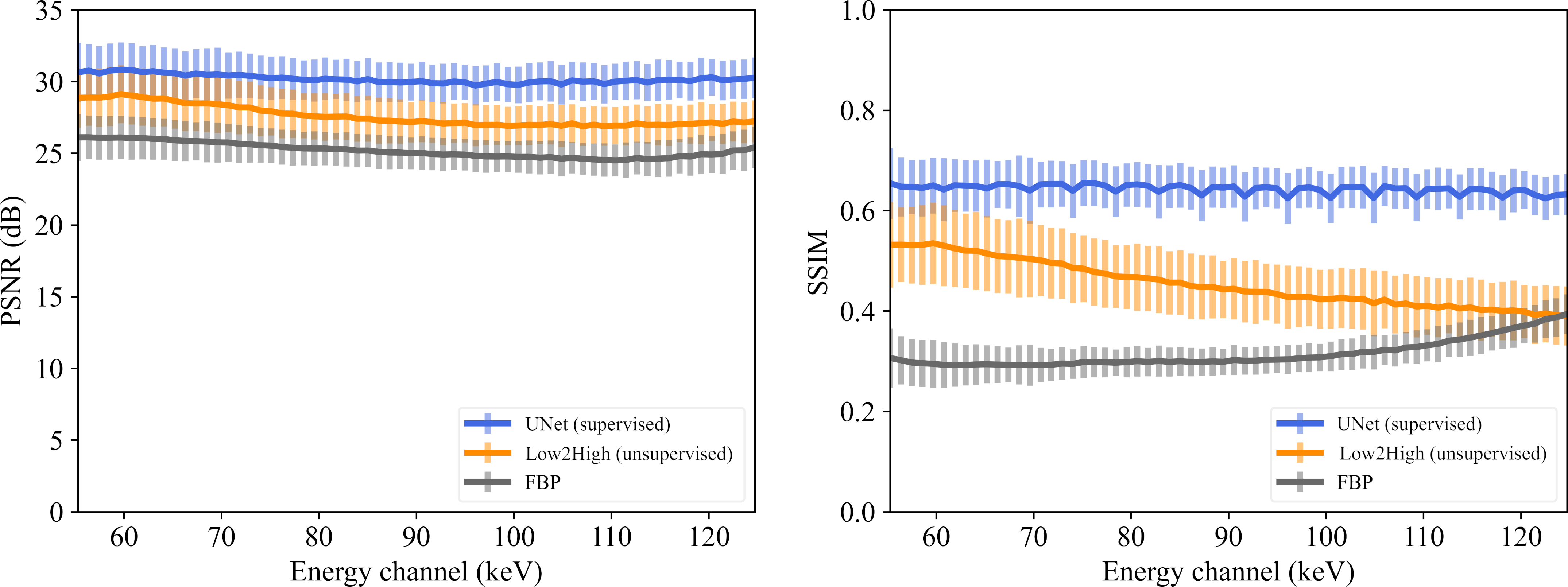}
\caption{Channel-wise PSNR and SSIM (mean $\pm$ SD) for spectral MUSIC dataset.  The highest PNSR and SSIM was observed with UNet.}
\label{fig:Chan_PSNR_SSIM_MUSIC}
\end{figure*}

\begin{figure*}[b!]
\centering
\includegraphics[width=170mm]{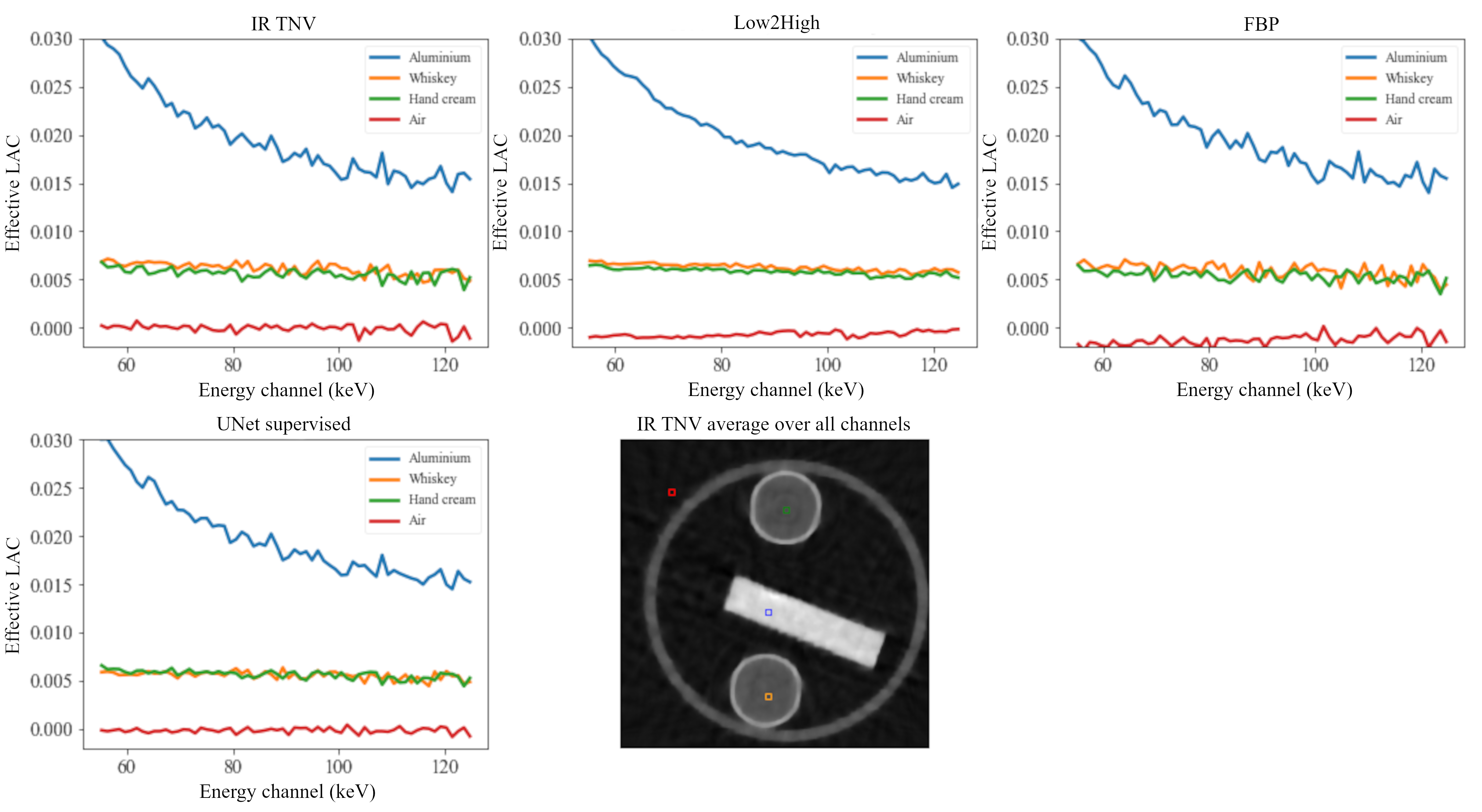}
\caption{Effective linear attenuation coefficients for different materials over energy channels for MUSIC dataset. Both, the leaning based methods and FBP were able to preserve the effective LAC values similarly as the reference IR TNV.}
\label{fig:LAC_channel_MUSIC}
\end{figure*}

\begin{figure*}[t]
\centering
\includegraphics[width=181mm]{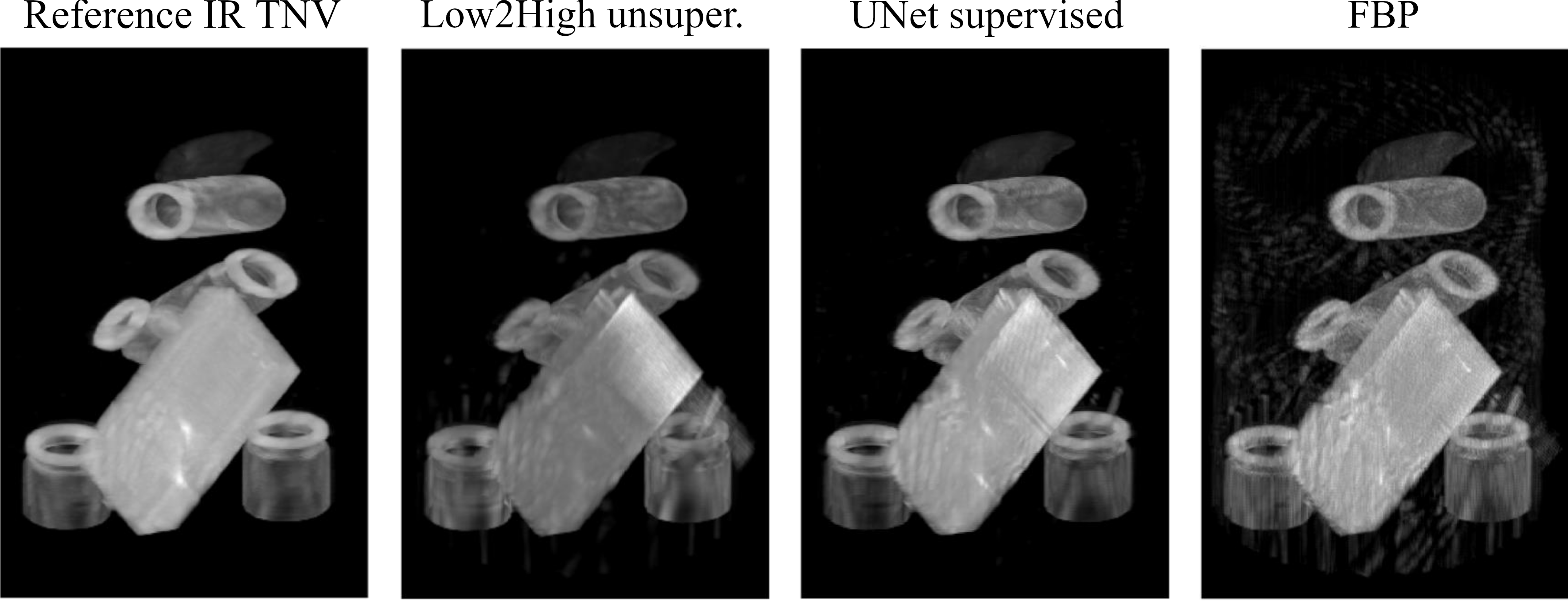}
\caption{Maximum intensity projection (MIP) images rendered from the test tube reconstructions (Sample 24012018) of MUSIC dataset. Streaking artefacts were observed for all MIP images at the lower end of the sample tube where there were aluminium bar and two test tubes.}
\label{fig:MIP_MUSIC}
\end{figure*}

When evaluated quantitatively, the supervised UNet denoiser had the most similar image quality performance to IR TNV (Table \ref{tab:table_IQ_MUSIC}).
However, the unsupervised Low2High had only slightly inferior performance in PSNR. The supervised UNet denoiser preserved the image quality similar to IR TNV over the spectral channels (Fig. \ref{fig:Chan_PSNR_SSIM_MUSIC}). We note, that these values are limitedly useful as the reference reconstructions are not perfect, especially for the high noise channels. In particular, the decline in SSIM for the unsupervised Low2High approach is due to a better noise suppression compared to the supervised training. \rev{The performance of Low2High and supervised UNet were similar when the CNR and SNR were assessed, see Fig. \ref{fig:ROI-Locations} for the target and Table \ref{tab:table_MUSIC_qualitative_IQ} for the results. Interestingly, both methods outperformed the reference IR TNV method, which again suggests a regularizing effect of training over all spectral channels. 
The Low2High method had a slightly higher CNR and SNR in low contrast region and UNet on the other hand had higher CNR and SNR in high contrast region.}

The effective LAC values were also preserved with the measured MUSIC dataset over the spectral channels for all reconstruction methods (Fig. \ref{fig:LAC_channel_MUSIC}) and most notable the Low2High method produced the smoothest value, i.e., least noise present in the LAC values. Finally, MIP images illustrated that each reconstruction method had streaking artefacts at the lower end of the aluminium bar (Fig. \ref{fig:MIP_MUSIC}). Both learning based methods had less streaking artefacts present in MIP images when compared with FBP.

\section{Discussion}\label{sec:discussion}
To address the computational complexity of spectral multi-bin PCD-CT, we studied the applicability of learning-based reconstruction methods, i.e., denoising UNet in supervised and unsupervised learning domains, for fast sparse multi-bin spectral CT reconstruction. We were able to show with both simulated synthetic and measured data that the learned reconstruction methods preserve the image quality in the spectral channels while keeping the computation time feasible.

Previous multi-bin spectral PCD-CT studies have overcome computational challenges by reducing the number of channels or re-binning channels prior to applying iterative reconstruction \cite{Busi2019, Jumanazarov2020}. Dimension reduction methods have been proposed as well such as principal component analysis or non-negative matrix factorization \cite{Kheirabadi2017}. However, in spectroscopic X-ray CT applications, it is beneficial that energy channels are preserved as the multi-bin information may enable a material separation to materials with similar attenuation properties and improve the accuracy of material decomposition depending on the decomposition method. Furthermore,  improved the material decomposition accuracy for high k-edge contrast agents gadolinium and iodine has been reported with increasing the number of energy channels \cite{Yveborg2015}.

In this study, we chose previously introduced UNet \cite{Jin2017} as our backbone network architecture. The UNet denoises the input FBP reconstruction. UNet was robust to the varying noise levels of the different energy channels. Also, the application of iterative TNV as a reference reconstruction to replace a ground truth label did not deteriorate the learning of the networks for the measured MUSIC dataset even though IR TNV had varying noise characteristics in energy channels. In fact, by training over all spectral channels the learned reconstruction produced qualitatively better reconstructions than the reference method.

The primary limitation within this study is given by  the quality of the training data, as this effectively limits the performance of the learning-based reconstruction methods. We note, that the quality of the measured MUSIC dataset was limited, since the projection data were acquired from 37 projection angles yielding a sparse-view problem, and data from only seven sample tubes were available. However, this resulted in 929 slice images with each 64 spectral channels for the training set for the supervised approach, which should be a feasible amount for the training of the model. 
The sparseness and noisiness of the MUSIC dataset led to the selection of iterative reconstruction with total nuclear variation as reference reconstruction to replace a ground truth label. For supervised training, one would preferably need separate measurements from densely sampled projection angle increments with higher exposure (i.e. less noise) to generate high-quality labels. Interestingly, the supervised method showed that when trained on a reference reconstruction it was capable to overcome a possible bias spectral bias by training over all channels simultaneously.

Given the limitations of the supervised approach, we have proposed an unsupervised training regime that is designed to work specifically for sparse spectral data. This is achieved by creating training pairs from two differently filtered backprojections, Hann and Ram-Lak. We combined the training pairs with a TNV penalty to suppress streaking artefacts. While this Low2High approach produces excellent reconstructions for the noisy high energy channels with a successful suppression of streaking artefacts the performance for low energy channels was not as satisfactory, where stronger residual streaking artefacts were visible.  Nevertheless, in comparison to the supervised method the proposed unsupervised Low2High training produced visually and quantitatively (by smoother LAC values) better reconstructions than the comparison methods for the experimental MUSIC data. In the simulated case, streaking artefacts where still more dominant in comparison.

Future studies should focus on expanding the learned spectral reconstruction methods to the three-dimensional spatial domain, which constitutes still a major computational burden due to large memory requirements and computation times limiting applications where fast reconstruction is needed e.g. in the security domain. In addition, a study addressing the further applicability of the reconstructed spectral slice images from the learned reconstructions e.g., in the estimation of electron density and effective atomic numbers as shown in \cite{Jumanazarov2020} warrants for investigation. Finally, a combination of the supervised and unsupervised method should be investigated to be enable training on a designated training set, instead of the test set, and allow for rapid application of the learned reconstruction method.

\section{Conclusion} \label{sec:conclusions}
We have presented an unsupervised reconstruction approach with total nuclear norm for sparse-view spectral multi-bin PCD-CT in comparison to a fully supervised variant. Both unsupervised and supervised UNet networks preserved the image quality and effective LAC values over the spectral channels in reconstructions. To conclude, the computation time of sparse angle multi-bin spectral CT reconstruction can be successfully reduced with learned reconstructions in a supervised manner while producing comparable image quality with even better noise suppression than a reference iterative reconstruction. If trained unsupervised additional regularization by a spectral coupling is needed. Reconstruction times are governed by the training procedure, but excellent results can be obtained for experimental measurement data, especially for noisy high energy channels.

\rev{Finally, we believe that the employed TNV penalty in the training procedure will also prove useful for other supervised approaches, such as DIP \cite{ulyanov2018deep,baguer2020computed}.}

\section*{Acknowledgment}
The authors would like to thank Allard Hendriksen for helpful discussions on the Noise2Inverse.
The authors wish to acknowledge CSC – IT Center for Science, Finland, for computational resources.

\bibliographystyle{IEEEtran}
\bibliography{library_v2,DLbib}

\end{document}